\documentclass{Rinton-P9x6}

\begin{document}

\title{Pseudo-Digital Qubits:  A General Approach}

\author{Mark Friesen, Robert Joynt, and M. A. Eriksson}

\address{Physics Department, University of Wisconsin-Madison, Madison, WI 53706, USA\\
E-mail: friesen@cae.wisc.edu, maeriksson@facstaff.wisc.edu}  

\maketitle

\abstracts{
Pseudo-digital coupling has recently been proposed as a simple but robust technique for reducing gating errors in quantum dot quantum computers.  
Here, we discuss the technique in the context of simulations on silicon heterostructures.
Additionally, we generalize and extend the pseudo-digital concept to other settings.
In particular, we consider superconducting charge qubits and suggest a simple circuit for implementing dual, pseudo-digital working points.}

Quantum computers are essentially analog devices.  
As such, errors can easily accumulate, to the point that calculations are ruined.
Fortunately, large-scale computations are still feasible, due to the development of quantum error correcting codes.\cite{preskill98}
Nevertheless, error management remains a primary concern in the development of quantum hardware.\cite{levy01}
In addition to environmental influences, errors can also arise due to the imperfect implementation of quantum gates. 
Such errors are more likely at high clock speeds.
As we shall show, hardware-based error suppression becomes the key for treating speed-induced gating errors.

Scalable devices are especially susceptible to gating errors, since large-scale applications typically need to run very fast.
In this work we focus on two important scalable systems.  
First, we review our previous results on semiconductor quantum dot devices.  
Then we extend our results to superconducting qubits. 
Semiconductors are particularly strong candidates for scalability, because of compatibility with established microelectronics technologies. 
In silicon-based devices, the decoherence properties are particularly good.\cite{loss98,kane98,vrijen00}
However operating speeds must be much greater than decoherence times; for electron spin qubits in Si quantum dots this still necessitates clock rates of order GHz.\cite{APL}
At this speed, the voltage pulses used to implement gate operations can only be administered with 1\% accuracy per gate, even using state-of-the-art electronics.\cite{APL}  
Such error rates are far above the fault-tolerance threshold.\cite{steane02}
To make matters worse, the response of the qubit exchange coupling $J$ to a gate voltage pulse $V$ is exponential in  conventional gating schemes (Fig.~1(a)), so that even small errors in $V$ induce large errors in $J$.
We would like to modify this conventional gating architecture to enable a more digital response function for $J(V)$, so that fluctuations in $V$ do not affect $J$, at least at special working points.  

In Ref.~[6] we described a hardware-based technique that provides such pseudo-digital working points, allowing two orders of magnitude improvements in gate error rates.  
In this scheme, $J(V)$ does not actually become flat (digital).
Instead, it attains a maximum at a special working point.
Fig.~1(c) gives a top-view of the improved, pseudo-digital gating architecture.
The gates are to be lithographically patterned over a silicon-germanium heterostructure as described in Ref.~[8].
In this design, the active layer is a strained silicon quantum well, containing a controlled number of electrons.
The top-gates provide lateral electrostatic confinement for forming quantum dots.
The structure of Fig.~1(c) defines two coupled quantum dots, each containing a single electron.
The dots are bistable, meaning that the electron can sit on either side of the dot, as controlled by the plunger gates.
The main idea of the design is that electrons move within their dots in parallel channels, as in Fig.~1(b).
When the electrons are at their closest approach, the exchange coupling $J$ attains a maximum, as
borne out in simulations (Fig.~1(d)).\cite{APL}  The maximum of $J(V)$ serves as the ``on" working point, while the exponentially small value of $J(V)$ when the qubits are well separated serves as the ``off" working point.

\begin{figure}[t]
\centerline{\epsfxsize=4.5in 
\epsfbox{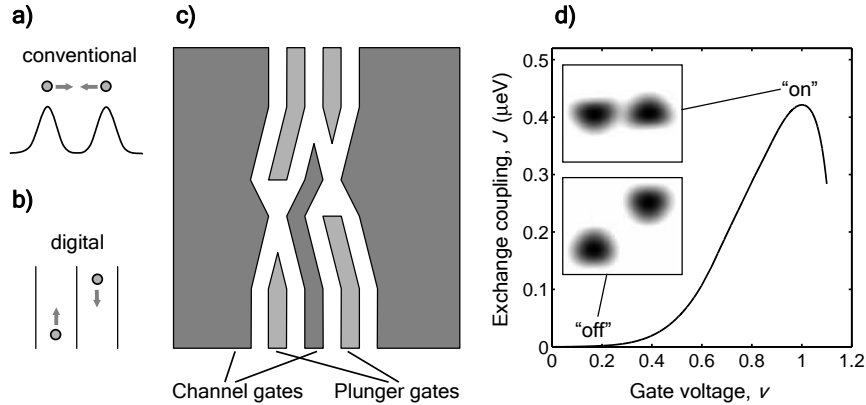}}
\caption{
Pseudo-digital coupling in quantum dots.  
(a) In the conventional gating scheme, a quantum barrier separates the two qubits.  
Lowering the barrier causes electrons to move towards each other on a line. 
Electronic exponential tails are responsible for the exponential dependence of $J(V)$. 
(b) In the pseudo-digital scheme, electrons move past each other in channels.  
(c) The top-gate structure used in the simulations (top-view).  
(d) The calculated exchange coupling $J$ as a function of plunger gate voltages $v$.  
(See Ref.~[6].)}
%\label{fig1}}
\end{figure}

The pseudo-digital control we have been discussing is very general, because its implementation requires only the presence of a maximum (or more generally, an extremum) in the qubit coupling $J$.  
To illustrate this point, the coupling $J$ between two quantum dot qubits can be expressed as a function of the relative coordinates $(x,y)$ of the two electrons (Fig.~2(a)).   
At first glance, the plot suggests no obvious extrema except at the unlikely working point $x=y=0$, corresponding to the complete overlap of the two electrons.  
However in the architecture of Fig.~1(c), electrons are made to move in channels, corresponding to a 1D trajectory in $(x,y)$ (Fig.~2(a), solid line).   
Constraining electron motion in this way allows us to define an arbitrary local maximum, or ``on" state.  
(We could similarly define a pseudo-digital ``off" state, with $J>0$, using a more complicated gate geometry:  Fig.~2(a), dashed line.)  

So far, we have discussed pseudo-digital coupling only in the context of interactions that depend on the physical positions of the qubits.  
Intriguingly, this technique applies equally well to qubit interactions where position plays no role.  
We consider a well known design for superconducting charge qubits (Fig.~3),\cite{makhlin01} made up of dc-SQUIDs threaded by individually controlled magnetic fluxes $\Phi_i$ and coupled by a common $LC$ oscillator mode.
The number of Cooper pairs on each lower island is tuned by separate voltage sources.
By varying the flux, the effective Josephson coupling can be tuned on each qubit, enabling single qubit operations.
Two qubit operations can also be accomplished by turning on the Josephson coupling in two qubits simultaneously. 
Under appropriate conditions,\cite{makhlin01} the qubit coupling strength reduces to $J\propto \cos (\pi \Phi_1/\Phi_0)\cos (\pi \Phi_2/\Phi_0)$, where $\Phi_0$ is the flux quantum.  
The parameters $\Phi_1$ and $\Phi_2$ control $J$, providing a natural 2D control basis for two-qubit operations (Fig.~2(b)), analogous to the $(x,y)$ basis used for quantum dots (Fig~2(a)).

\begin{figure}[t]
\centerline{\epsfxsize=4in 
\epsfbox{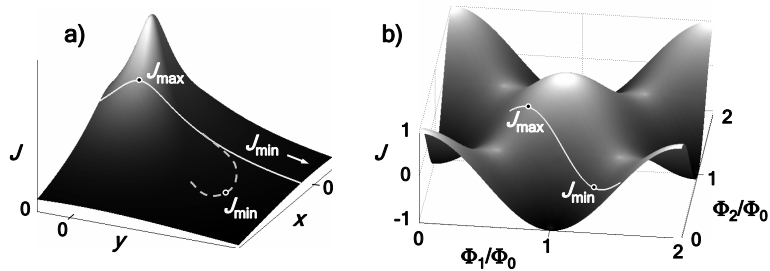}}
\caption{
1D coupling trajectories (white lines) plotted on 2D coupling parameter space.  
(a) Quantum dots:  $x$ and $y$ reflect the relative positions of the qubits, controlling the exchange coupling $J$.  
(b)  Superconducting charge qubits:  $\Phi_1$ and $\Phi_2$ reflect the magnetic flux through the two qubits, controlling the Josephson coupling.}
\end{figure}

To implement a pseudo-digital coupling for the charge qubit, we now need to define a 1D trajectory through $(\Phi_1,\Phi_2)$.  
We desire the trajectory to exhibit a local maximum $J_{\rm max}$ as well as a local minimum $J_{\rm min}$.   
The ability to define arbitrary working points provides a means to compensate for device-to-device variations which may prevent working at the ideal working points.
For the technique to be effective, the qubits cannot stray from their trajectories. 
Some uncertainty {\em along} the trajectory can be tolerated however--this is the main concept behind pseudo-digital control.
We enforce the single-trajectory requirement using the simple circuit shown in Fig.~3.
A single current source $I$ parameterizes the trajectory between the desired working points.  
The regular shape of $J(\Phi_1,\Phi_2)$ in Fig.~2(b) makes it easy to identify an appropriate trajectory.  
In fact, a straight line in $(\Phi_1,\Phi_2)$ works well.  
The flux through the $i$th qubit is generated by an external current loop, $\Phi_i = L_iI+\phi_i$, where $L_i$ is the loop inductance and $\phi_i$ is a constant external flux bias.  
The desired linear trajectory $\Phi_1(I)=b\Phi_2(I)+c$ can be achieved by adjusting the circuit parameters such that $L_1=bL_2$ and $\phi_1=b\phi_2+c$.  
With this circuit, fast switching between the working points can be accomplished, without incurring significant errors.
The key is to use a {\em single} current source to enforce the qubit coupling trajectory.

We have shown how to extend the pseudo-digital coupling scheme to superconducting charge qubits, in addition to quantum dot qubits.  
However the principle is very general:  (1) identify a 2D coupling control space, (2) identify a suitable 1D trajectory for qubits in that control space, (3) use hardware to confine qubits to their trajectories.  
Indeed, many quantum settings possess the necessary degrees of freedom to implement pseudo-digital control.

\begin{figure}[t]
\centerline{\epsfxsize=1.25in 
\epsfbox{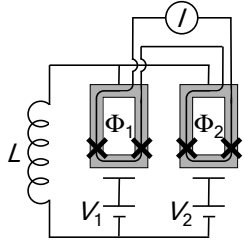}}
\caption{
Superconducting charge qubits, coupled through an $LC$ oscillator, and controlled via external fluxes $\Phi_1$ and $\Phi_2$ and gate voltages $V_1$ and $V_2$.$^9$
An additional current source is introduced to control $\Phi_1$ and $\Phi_2$.}
\end{figure}

\end{document}